\documentclass[letterpaper,superscriptaddress,twocolumn,showpacs,amsmath,amssymb,nofootinbib]{revtex4}

\makeatletter
\def\ignorecitefornumbering#1{%
     \begingroup
         \@fileswfalse
         #1
    \endgroup
}
\makeatother

\usepackage{tikz}
\usetikzlibrary{tikzmark}
\usepackage{graphicx,color,mathtools}
\usepackage[colorlinks,plainpages=false,linkcolor=black,urlcolor=blue,citecolor=blue,pdfpagemode=UseNone,pdfstartview=FitH]{hyperref}

\begin{document}

\title{Electronic band structure of optimal superconductors: \\from cuprates to ferropnictides and back again}

\author{A.~A.~Kordyuk}
\affiliation{Kyiv Academic University, 03142 Kyiv, Ukraine}
\affiliation{Institute of Metal Physics of National Academy of Sciences of Ukraine, 03142 Kyiv, Ukraine}

\begin{abstract}
While the beginning decade of the high-$T_c$ cuprates era passed under domination of local theories, Abrikosov was one of the few who took seriously the electronic band structure of cuprates, stressing the importance of an extended Van Hove singularity near the Fermi level. These ideas have not been widely accepted that time mainly because of a lack of experimental evidence for correlation between saddle point position and superconductivity. In this short contribution, based on the detailed comparison of the electronic band structures of different families of cuprates and iron based superconductors I argue that a general mechanism of the $T_c$ enhancement in all known high-$T_c$ superconductors is likely related with the proximity of certain Van Hove singularities to the Fermi level. While this mechanism remains to be fully understood, one may conclude that it is not related with the electron density of states but likely with some kind of resonances caused by a proximity of the Fermi surface to topological Lifshitz transition. One may also notice that the electronic correlations often shifts the electronic bands to optimal for superconductivity positions.
\end{abstract}
\pacs{74.20.-z, 74.25.Jb, 74.70.Xa, 74.72.-h, 79.60.–i}
\maketitle

\section{Introduction}

\begin{figure}[b]
\begin{center}
\includegraphics[width=0.38\textwidth]{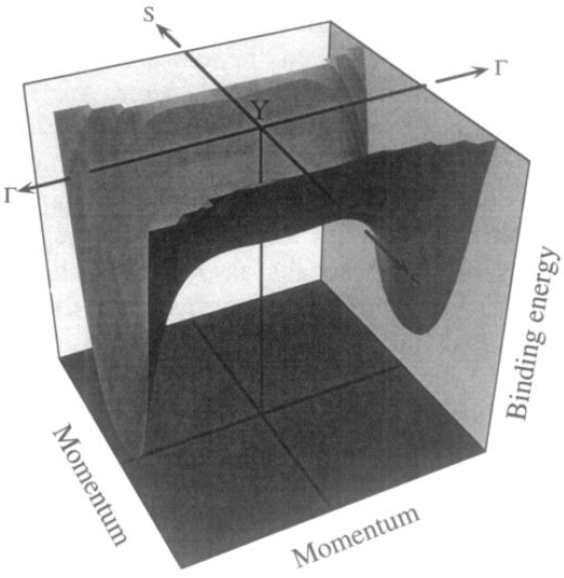}
\caption{The "extended saddle point" type Van Hove singularity derived from photoemission data from YBa$_2$Cu$_4$O$_8$: The momentum axes cover an interval of 1 \AA$^{-1}$ centered on $(\pi,0)$ point of the Brillouin zone, and the binding energy axis spans 0 to 60 meV \protect\ignorecitefornumbering{\cite{Gofron1994}}.
\label{VHS}}
\end{center}
\end{figure}

In spite of the fashion for the "local language" \cite{Anderson1987, Dagotto1994, Imada1998} in application to physics of high-$T_c$ cuprates soon after their discovery, some researchers were keeping to believe that it is the electronic band structure of cuprates that conceals a key to understand them \cite{Hirsch1986, Dzialoshinskii1987, Gorbatsevich1987, Labbe1987, Friedel1989, Markiewicz1989}. Especially many efforts had been made to study possible consequences of close vicinity of the "saddle point" type Van Hove singularity (VHS) \cite{VanHove1953} to the Fermi level (see Fig.\;\ref{VHS}), as it had been revealed by the band structure calculations \cite{Markiewicz1997} and earlier photoemission experiments in a number of cuprates \cite{Gofron1994, King1994, Yokoya1996}. Abrikosov, who advocated "common sense against fashion" in this respect, was fascinated by rich physics that comes from an "extended" saddle-point and had derived several formulas to describe it analytically \cite{Abrikosov1993, Gofron1994, Abrikosov1994, Abrikosov1998} suggesting finally his "theory of high-$T_c$ superconducting cuprates based on experimental evidence" \cite{Abrikosov1999, Abrikosov2000, Abrikosov2008}.

\begin{figure*}
\begin{center}
\includegraphics[width=1\textwidth]{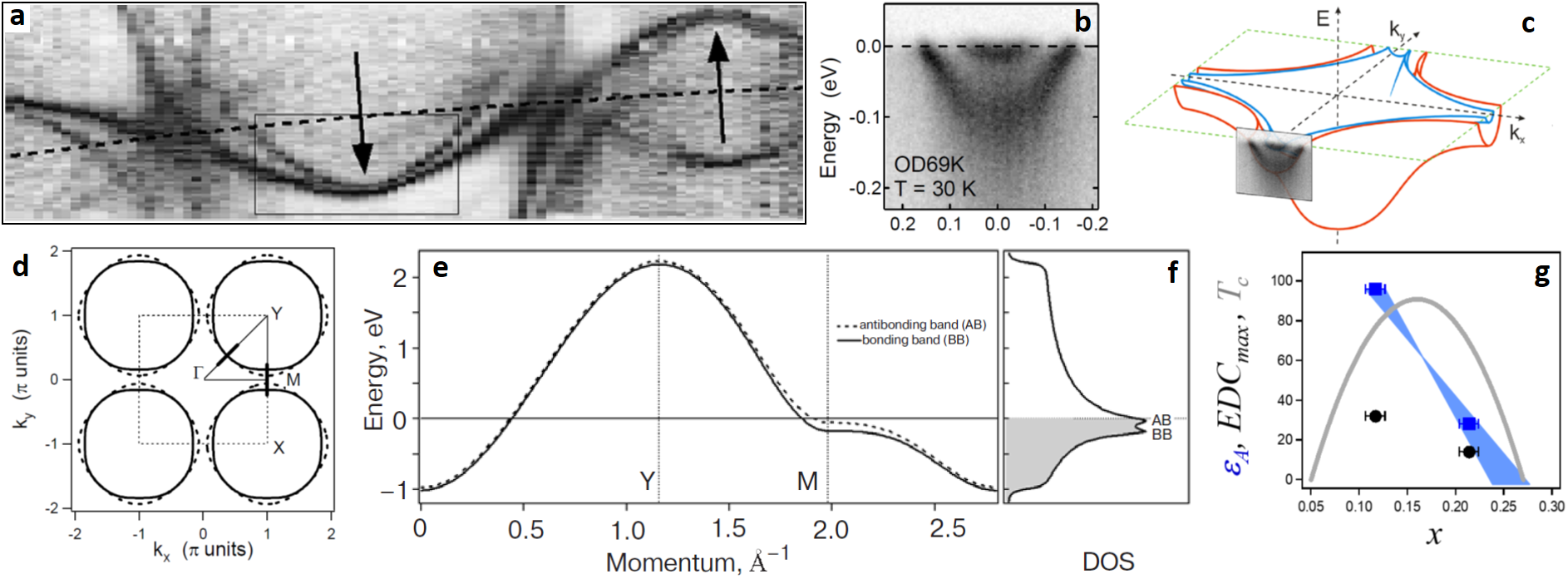}
\caption{Electronic band structure of an overdoped Bi-2212 derived from ARPES experiments: a fragment of the Fermi surface (arrows indicate the nodal FS crossings with well resolved bilayer splitting) (a)  \protect\ignorecitefornumbering{\cite{2004_PRB_Kordyuk}}; the ARPES spectrum represents the anti-nodal crossing of the saddle point (b)  \protect\ignorecitefornumbering{\cite{2003_PRL_Borisenko}}; the bare band structure (c), FS contours (d), dispersions (e) and DOS (f) derived from experimental data \protect\ignorecitefornumbering{\cite{2003_PRB_Kordyuk}}; the position of the saddle point of the antibonding band derived from the data (EDC maximum, black circles) and bare band positions ($\varepsilon_A$, blue squares) in meV are shown together with $T_c$ (grey solid line) in K as function of hole concentration (g) \protect\ignorecitefornumbering{\cite{2006_LTP_Kordyuk}}.
\label{Band_structure}}
\end{center}
\end{figure*}

The role of saddle point VHS has been discussed in two types of scenarios. The "direct" scenarios relate the $T_c$ enhancement with VHS related peak in the density of states (DOS) \cite{Labbe1987, Xu1987, Markiewicz1989}, which for the "extended" singularity \cite{Friedel1989, Gofron1994} leads to the stronger than logarithmic divergence ("a power law divergence") in DOS \cite{Abrikosov1993, Abrikosov2000}. In the "indirect" scenarios, the superconductivity is enhanced by competing instabilities \cite{Dzialoshinskii1987, Gorbatsevich1987, Hirsch1986, Markiewicz1991, Kampf2003}. Moreoover, it has been found that strong correlation effects pin this VHS close to the Fermi level \cite{Markiewicz1990, Newns1991, Si1993, Monthoux1993, Andersen1994, Liechtenstein1996}. Other aspects of the saddle point VHS, like the dynamic VHS-Jahn-Teller effect, the pseudogap and striped phases, are discussed in detail in another review by Markiewicz \cite{Markiewicz1997a}. 

The discussed singularity in DOS was later shown to be rather weak to account for high $T_c$'s, especially when finite temperature and impurity scattering are taken into account \cite{Pashitskii2006, Plakida2010}. Applicability of the models with competing instabilities is more difficult to estimate but the overall frustration about them have been arisen mainly because of a lack of experimental evidence for correlation between the position of the saddle point and superconductivity: a number of cuprates with VHS close to the Fermi level show rather low $T_c$'s \cite{Markiewicz1997, King1994, Yokoya1996}. In addition, while the saddle point stays close to the Fermi level for the hole doped cuprates, it goes deeper with hole underdoping and should continue to sink further with the electron doping. So, any scenarios of $T_c$ enhancement related to the saddle point VHS cannot universally explain the both sides of the electronic phase diagram.  

Here, based on overview of a number of photoemission data, I argue that indeed there is a robust correlation between superconducting critical temperature and a proximity of certain Van Hove singularities to the Fermi level in all known high-$T_c$ superconductors including the iron based superconductors (Fe-SC) and high-$T_c$ cuprates (Cu-SC) on the both sides of the phase diagram. Interestingly, this VHS is usually not a saddle point but an edge (top or bottom) of certain bands which, in the vicinity to topological Lifshitz transition \cite{Lifshitz1960} plays, most likely, a role of a "resonant amplifier" of superconductivity in a multi-band system. While we are looking for microscopic understanding of this correlation, it can be used to search for new high temperature superconductors with higher $T_c$'s.

\begin{figure*}
\begin{center}
\includegraphics[width=0.9\textwidth]{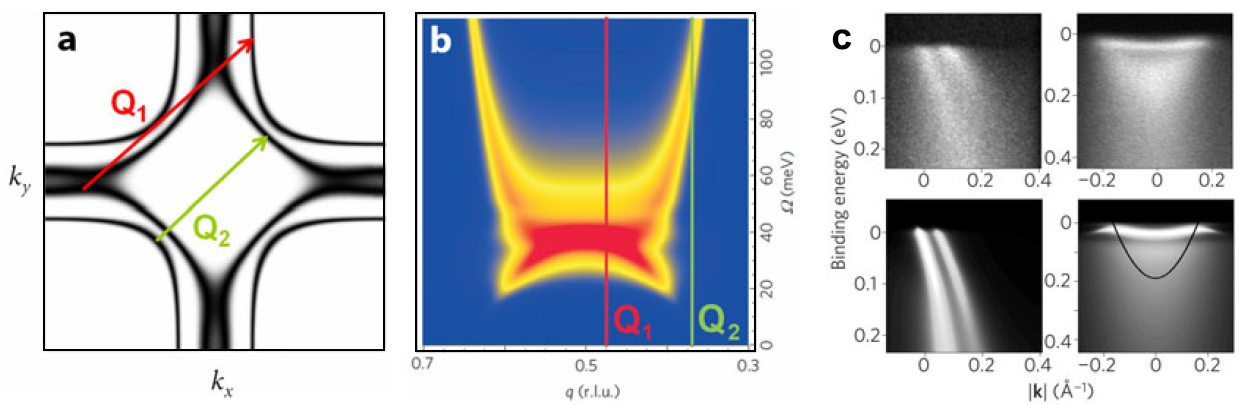}
\caption{"Fingerprints" of the spin-fluctuations in ARPES spectra of cuprates. (a) The Fermi surface of YBCO in the 1st Brillouin zone derived from ARPES data represents the fermionic Green’s function. (b) The spin excitations spectrum resulting from numerical fits to the inelastic neutron scattering data. (c) Comparison of experimental (upper row) and theoretical (lower row) fermionic spectra \protect\ignorecitefornumbering{\cite{2009_NP_Dahm,2010_EPJST_Kordyuk}}.
\label{ARPES_INS}}
\end{center}
\end{figure*}

\section{Electronic band structure}
\subsection{Cuprates}

As it has been mentioned, when Abrikosov was exploring the consequences of the extended saddle point in the electronic band structure of the cuprates, most of the researchers did not believe that the concepts of the one-particle electronic structure or the Fermi liquid are applicable to the cuprates at all. However, already the first angle resolved photoemission experiments on Bi$_2$Sr$_2$CaCu$_2$O$_{8+x}$ (BSCCO or Bi-2212) \cite{Takahashi1988, Dessau1993}, YBa$_2$Cu$_3$O$_{7-\delta}$ (YBCO) \cite{Liu1992}, and Nd$_{2-x}$Ce$_x$CuO$_4$ \cite{King1993} had revealed the dispersion and the Fermi surface very similar to those obtained by conventional density-functional band-structure calculations. The essential development of angle resolved photoemission spectroscopy (ARPES) during the next decade has allowed to shed much more light on this issue revealing the details of the band structure and quasiparticle spectrum of cuprates \cite{Damascelli2003, 2014_LTP_Kordyuk}.

In particular, it has been shown that the hole doped cuprates have a large Fermi surface (FS) in the range of doping when they are superconducting \cite{Aebi1994, Borisenko2000, 2001_PRB_Borisenko, 2002_PRB_Kordyuk}. This FS satisfies the Luttinger theorem, i.e., its volume corresponds to the number of the conduction electrons per unit cell and is proportional to $(1 - x)$, where $x$ is the hole concentration. Then \cite{Feng2001, Chuang2001, 2002_PRB_Kordyuk, 2002_PRL_Kordyuk, 2004_PRB_Kordyuk}, most of ARPES-groups began to observe the splitting of the conduction band in the bilayer cuprates into the sheets corresponding to the bonding and anti-bonding orbitals (see Fig.\;\ref{Band_structure}) that contradicted the idea of spatial confinement of electrons in separate layers \cite{Anderson1987, Ding1996}. Note that most of ARPES results for cuprates have been obtained on Bi-2212, since the bulk properties of Y-123 are much more difficult to study because of overdoped non-superconducting topmost layer \cite{2007_PRB_Zabolotnyyb, 2007_PRB_Zabolotnyy}.

Based on the FS geometry and low energy electron dispersions one may derive the hopping integrals describing the conduction band \cite{2003_PRB_Kordyuk} and, for two-CuO$_2$-layer Bi-2212 we have derived that the onset of the superconducting region in the phase diagram, in the direction of reducing the hole concentration, starts with the Lifshitz topological transition for the anti-bonding Fermi surface \cite{2006_LTP_Kordyuk} (Fig.\;\ref{Band_structure}g). While the same holds for a single FS of single layer Bi-2201 \cite{Kondo2004}, a careful study of Bi-2212 of different doping levels \cite{Kaminski2006} has shown that the Lifshitz transition for the anti-bonding band appears a bit later, at between 0.22 and 0.23 holes per Cu atom ($T_c =$ 55 K), that is similar to earlier result on ${\mathrm{La}}_{2\ensuremath{-}x}{\mathrm{Sr}}_{x}{\mathrm{CuO}}_{4}$ \cite{Ino2002}.  

Another conclusion that has been derived from the numerous ARPES experiments is that the whole spectra of the superconducting cuprates (except may be the pseudogap effect) can be described by the quasiparticle spectral function \cite{2006_LTP_Kordyuk, 2010_EPJST_Kordyuk}, $A \propto$ Im($G$), $G^{-1} = G_0^{-1} - \Sigma$, in which the bare Green's function $G_0 = 1/(\varepsilon_k - \omega + i0)$ with the bare electron dispersion $\varepsilon_k$ are defined by the interaction of the electrons with periodic crystal lattice and the quasiparticle self-energy $\Sigma$ encapsulates the interaction of electron with other electrons and other degrees of freedom, like in normal metals \cite{Valla1999}. Yet the striking difference of ARPES spectra of the cuprates from the spectra of normal metals is in strong scattering that (1) does not stop at the Debye energy (so, the dispersion is hard to follow below $-0.3$ eV \cite{2005_PRB_Kordyuk, 2007_PRL_Inosov}) and (2) is strongly momentum dependent, leading to the "nodal-antinodal dichotomy": around $(\pi,0)$ point of the Brillouin zone (the "antinodal" region) the renormalization is highly increasing below $T_c$, while along the nodal direction its temperature dependence is rather weak \cite{2003_PRL_Borisenko, 2003_PRL_Kim, 2006_LTP_Kordyuk}.

The analysis of the self-energy dependence on energy, momentum \cite{Kaminski2001, 2003_PRL_Kim}, temperature and doping level \cite{2004_PRL_Kordyuk, 2005_PRB_Kordyuk, 2006_PRL_Kordyuk} has indicated that the main channel of one-electron excitation scattering is related with the spin-fluctuations. The direct comparison of the ARPES and inelastic neutron scattering spectra (Fig.\;\ref{ARPES_INS}) has proved this idea \cite{2009_NP_Dahm}, naturally resolving the nodal-antinodal dichotomy: the self-energy of the nodal quasiparticles is defined by scattering by high energy branches of the spin-fluctuation spectrum while the antinodal self-energy is formed by the scattering by the magnetic resonance \cite{Eschrig2006} formed below $T_c$. We can consider $\Sigma$ as a generalized cross-correlation \cite{2010_EPJST_Kordyuk} of one-particle spectrum presented by the Green's function and the two-particle spectrum of spin-fluctuations: $\Sigma = \bar{U}^2 \chi \star G$, where $\bar{U}$ is taken for a spin-electron coupling constant. Also, it has been shown \cite{Chatterjee2007, 2007_PRB_Inosov} that the spin-fluctuation spectrum itself is formed by itinerant electrons: $\chi = G \star G$. So, one can write an extended Dyson equation for Cu-SC (see \cite{2010_EPJST_Kordyuk} for details):
\begin{equation}\label{E1}
G^{-1} = G^{-1}_0 - \bar{U}^2 G \star G \star G.
\end{equation}
     
Therefore, one could conclude that "a conservative view" of Friedel \cite{Friedel1989} in late 80's, that treating the on-site electron interactions as a perturbation to a band scheme is sufficient to describe the physical properties of superconducting cuprates, is largely supported by later ARPES experiments. 

The issue of the pseudogap in cuprates is rather complicated since evidently encapsulates a number of mechanisms \cite{2015_LTP_Kordyuk}, some of which, like charge or spin density waves can be described by an effect of the new order, but still there is a place for localization effects. I will turn back the the pseudogap issue in Section \ref{SecPG}.    

To summarize, the electronic band structure (ES) of cuprates defines the spin-fluctuation spectrum (SF) and the electronic ordering, which likely forms the pseudogap (PG) state. An interplay of ES with SF leads to superconductivity (SC) that competes with PG. So, one can write one more formula for Cu-SC: 
\vspace{6pt}
\begin{equation}\label{E2}
\text{E\tikzmark{0}\tikzmark{1}S} + \text{S\tikzmark{00}F} - \text{P\tikzmark{11}G} \Rightarrow 
 \text{SC}. \\[8pt]
\end{equation}
The pairing by the spin-fluctuations can lead to high $T_c$ in some theories \cite{Monthoux1994, Abanov2002, Maier2007} or cannot in point of view of others \cite{Kee2002, Alexandrov2007}.
\begin{tikzpicture}[remember picture, overlay, bend left=45, -latex, black]
  \draw ([yshift=2.3ex]pic cs:0) to ([yshift=2.2ex]pic cs:00);
\end{tikzpicture}
\begin{tikzpicture}[remember picture, overlay, bend left=-40, -latex, black]
  \draw ([yshift=-0.5ex]pic cs:1) to ([yshift=-0.2ex]pic cs:11);
\end{tikzpicture}

\begin{figure*}
\begin{center}
\includegraphics[width=1\textwidth]{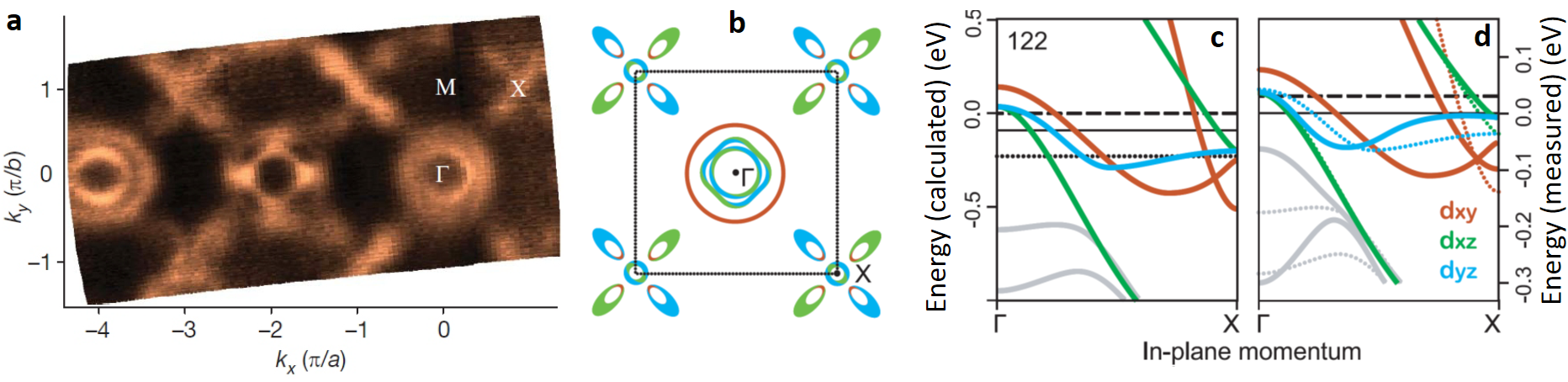}
\caption{Electronic structure of iron based superconductors (Fe-SC): (a) Fermi surface (FS) maps measured by ARPES for an optimally doped Ba$_{1-x}$K$_{x}$Fe$_2$As$_2$ (BKFA) \protect\ignorecitefornumbering{\cite{2009_N_Zabolotnyy}}; (b) the scketch of FS derived from comparison of experimental spectra to band structure calculations with indication (by color) of the orbital characted of corresponding bands; the comparison of the calculated (c, and dotted lined on d) and experimental (d) band structure along the $\Gamma$X direction of the Brillouin zone \protect\ignorecitefornumbering{\cite{2012_LTP_Kordyuk, 2013_JSNM_Kordyuk}}.
\label{FeSC}}
\end{center}
\end{figure*}

\subsection{Iron based superconductors}

The band structure of the iron based superconductors (Fe-SC) is much more complex than of Cu-SC and consists usually of five conduction bands crossing the Fermi level \cite{2012_LTP_Kordyuk, 2013_JSNM_Kordyuk} (see Fig.\;\ref{FeSC}). It is well captured by DFT calculations \cite{Andersen2011, Sadovskii2012} but do not take it too literally. The calculated Fermi surface is usually bad starting point for theory, since even topology of the Fermi surface is very sensitive to slight shifts of the bands in respect to $E_F$ and to each other and often differs in experiment and calculations \cite{2009_N_Zabolotnyy, 2010_PRL_Borisenko}. 

The band structures of Fe-SC seen in the ARPES experiments differ from the calculated ones mainly in two ways: a strong renormalization (3 times in avarage \cite{2010_PRL_Borisenko, 2011_PRB_Kordyuk} but band-dependent \cite{Maletz2014, Fanfarillo2016} and peaked at about 0.5 eV \cite{2017_PRB_Evtushinsky}), and a momentum-dependent shift \cite{Yi2009, 2010_PRL_Borisenko, Brouet2013, Watson2015} (the "red-blue shift" \cite{2016_LTP_Pustovit, Lochner2017}).

The bands forming the Fermi surfaces of Fe-SC have distinct orbital characters mainly of three types: Fe $3d_{xy}, 3d_{xz}, and 3d_{yz}$ (Fig.\;\ref{FeSC} b-d). Moreover, it is the $d_{xz}/d_{yz}$ bands that carry the largest superconducting gap \cite{Ding2008, 2009_NJP_Evtushinsky, 2012_S_Borisenko, 2014_PRB_Evtushinsky} and are therefore the most important for superconductivity in Fe-SC. This simplifies the situation a bit, but, on the other side, the AF ordered phase and preceeding nematic transition \cite{Fernandes2014} essentially complicates the electronic band structure of the "normal" state from which the superconductivity occurs. 

From the theory side, the question what drives both the supercoducting pairing and the nematic ordering remains open. The phonons alone, despite some "firgerprints" they left in the ARPES spectra \cite{2011_PRB_Kordyuk}, are not considered seriously \cite{Hirschfeld2011, Fernandes2014}. Even the state is called "nematic" rather than "anisotropic" to stress the electronic origin of the instability, and there is a lot of experimental evidences for this \cite{Fernandes2014}, but one prefers to speak about interplay of phonons, charge/orbital fluctuations, and spin fluctuations \cite{Yin2010, Onari2012, Chubukov2012, Fernandes2014, Chubukov2016}. Although it is agreed that phonons and charge/orbital fluctuations would favour a sign-preserving $s$-wave superconducting order parameter ($s^{++}$) whereas spin-fluctuations favour a sign-changing $s$-wave ($s^{+-}$) superconductivity \cite{Wang2011, Hirschfeld2011, Fernandes2014}, any agreement on why $T_c$'s are so high is absent so far and there is no confirmed prediction for new high temperature superconductors.

\section{Lifshitz transition}

\subsection{Iron based superconductors}

\begin{figure*}
\begin{center}
\includegraphics[width=0.9\textwidth]{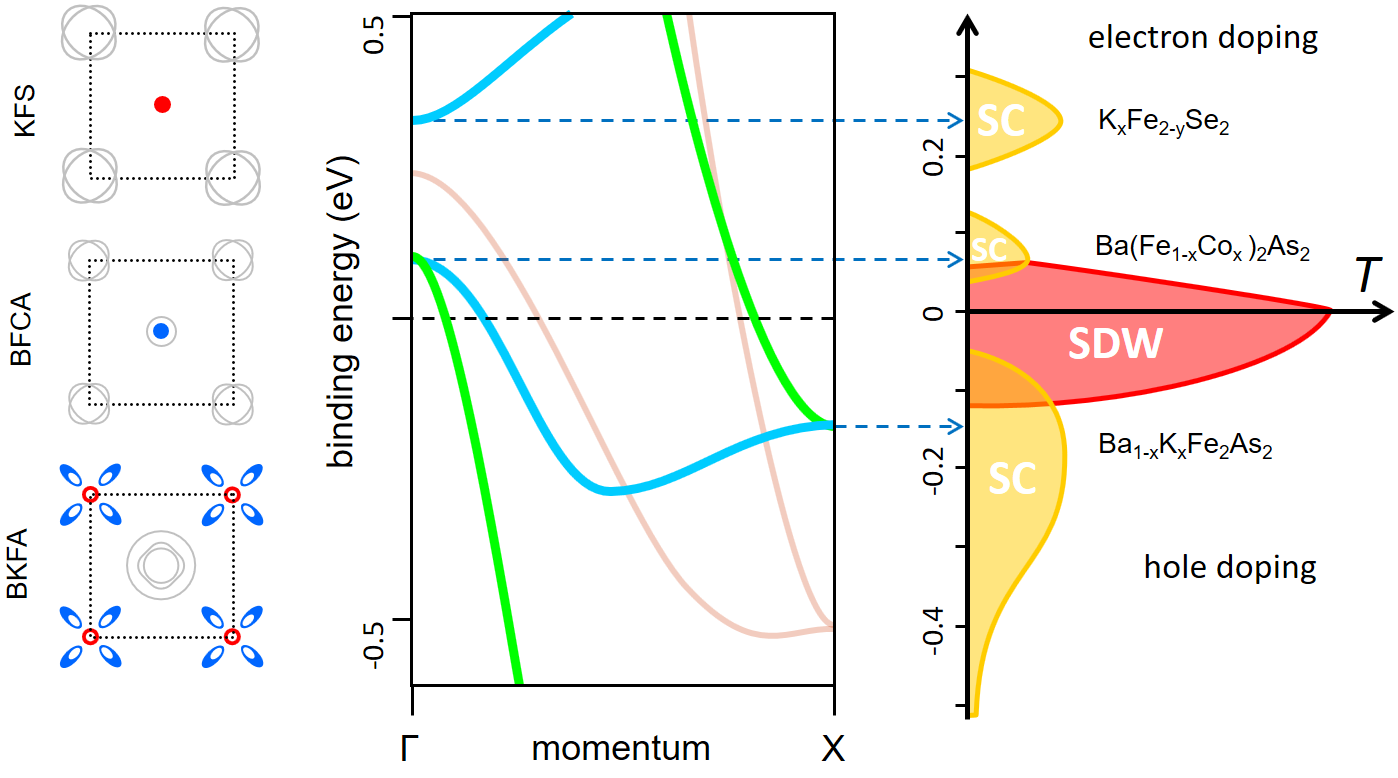}
\caption{The LT-$T_c$ correlation \protect\ignorecitefornumbering{\cite{2012_LTP_Kordyuk, 2013_JSNM_Kordyuk}} is illustrated through a projection of the Fermi level crossing the "rigid" electronic band structure of Fe-SC (central panel) on the charge carrier concentration scale of the phase diagram (right): $T_c$ maxima correspond to proximity of the tops/bottoms of $d_{xz}/d_{yz}$ bands to the Fermi level; the corresponding "optimal" Fermi surfaces (blue for hole- and red for electron-like sheets) are shown on the left. 
\label{PhD_FeSC}}
\end{center}
\end{figure*}

From the experiment side, the complexity of the band structure of Fe-SC seems to play a positive role in the struggle for understanding the pairing mechanism because the multiple electronic bands and the resulting complex fermiology offer exceptionally rich playground for establishing useful empirical correlations. In particular, there is an empirical correlation between the electronic structure and $T_c$: maximal $T_c$ (optimally doped superconductors) is observed when a proximity of the electronic structure to topological Lifshitz transition \cite{Lifshitz1960} takes place \cite{2012_LTP_Kordyuk, 2013_JSNM_Kordyuk}. Interestingly, this Lifshitz transition (LT) is related with VHS wich is usually not a saddle point but an edge (top or bottom) of certain bands, namely $d_{xz}/d_{yz}$ bands, as shown in Fig.\;\ref{PhD_FeSC}.

This LT-$T_c$ correaltion is observed for all known optimally doped Fe-SC except, may be, some FeSe-based compounds \cite{Lee2014, 2016_LTP_Pustovit}. Indeed, the extremely small Fermi surface sheets of $d_{xz}/d_{yz}$ orbital origin are observed for the optimally hole doped Ba$_{1-x}$K$_{x}$Fe$_2$As$_2$ (BKFA) \cite{2009_N_Zabolotnyy, 2009_NJP_Evtushinsky, 2011_JPSJ_Evtushinsky}, Ba$_{1-x}$Na$_{x}$Fe$_2$As$_2$ (BNFA) \cite{2012_PRB_Aswartham}, and Ca$_{1-x}$Na$_{x}$Fe$_2$As$_2$ \cite{2013_PRB_Evtushinsky} as well as for the optimally electron doped Ba(Fe$_{1-x}$Co$_{x}$)$_2$As$_2$ (BFCA) \cite{Liu2011, 2013_JSNM_Kordyuk}, i.e. for the both sides of the electron phase diagram for the 122 system. The same holds for the stoichiometric (but optimal for $T_c$) LiFeAs \cite{2010_PRL_Borisenko, 2011_PRB_Kordyuk, 2012_S_Borisenko},  NaFeAs \cite{He2010, 2012_PRB_Thirupathaiah}, and for A$_x$Fe$_{2-y}$Se$_2$ family (A stands for alkali metal: K, Rb, Cs, and Tl) \cite{Mou2011, Zhang2011, 2013_PRB_Maletz}. Moreover, the $T_c$ is increasing with the number of band-edge VHS's at $E_F$, as it has been shown comparing (CaFeAs)${}_{10}$Pt${}_{3.58}$As${}_{8}$ (three band-edge VHS's at $E_F$, $T_c$ = 35 K) to (CaFe${}_{0.95}$Pt${}_{0.05}$As)${}_{10}$Pt${}_{3}$As${}_{8}$ (only one VHS at $E_F$, $T_c$ = 15 K) \cite{Thirupathaiah2013}. Finally, now one can say that the same is true for the 1111-type compounds which exhibit the highest $T_c$ up to 55 K. Having the polar surfaces, these compounds are hard to study by ARPES \cite{2012_LTP_Kordyuk, 2014_LTP_Kordyuk}, but it has been shown that the bulk electronic structure for SmFe$_{0.92}$Co$_{0.08}$AsO \cite{Charnukha2015} and NdFeAsO$_{0.6}$F$_{0.4}$ ($T_c$ = 38 K) \cite{Charnukha2015a} is in the same optimal for superconductivity state, having 2-3 band-edge VHS's in close vicinity to the Fermi level. 

As for FeSe, its Fermi surfaces look a bit away from the Lifshitz transition \cite{2016_LTP_Pustovit}, but pure FeSe crystals are not optimal for superconductivity: their $T_c$ increases from about 9 to 38 K under pressure \cite{Medvedev2009} and by means of intercalation \cite{Guo2010}. While it is hard to do ARPES under such a pressure, the results of a DFT+DMFT calculations show that the bulk FeSe under pressure about 9 GPa undergoes a Lifshitz transition \cite{Skornyakov2018}.
 
The possible mechanism of this correlation will be briefly discussed in Section \ref{RSC}, but it would be tempting to use the observed correlation for a search of new high temperature superconductors with higher $T_c$'s. Similar electronic band structure for all the Fe-SC's results in similar DOS \cite{2013_JSNM_Kordyuk} from which one can clearly see that this correlation has nothing to do with DOS enhancement. The bright example is KFe$_2$As$_2$ that has much higher DOS at $E_F$ than any of optimally doped Fe-SC's and $T_c$ about 4 K. On the other hand, the density of states should be certainly important for superconductivity, so, looking for Fe-SC's with higher $T_c$ one should find a compound with several bands crossing the Fermi level one or more of which are close to the Lifshitz transition but with the higher density of states from the other bands (similarly to hole-doped 122). This should be the case for hole overdoped KFe$_2$As$_2$ or LiFeAs.

\subsection{Cuprates}
\label{SecPG}

\begin{figure*}
\begin{center}
\includegraphics[width=0.8\textwidth]{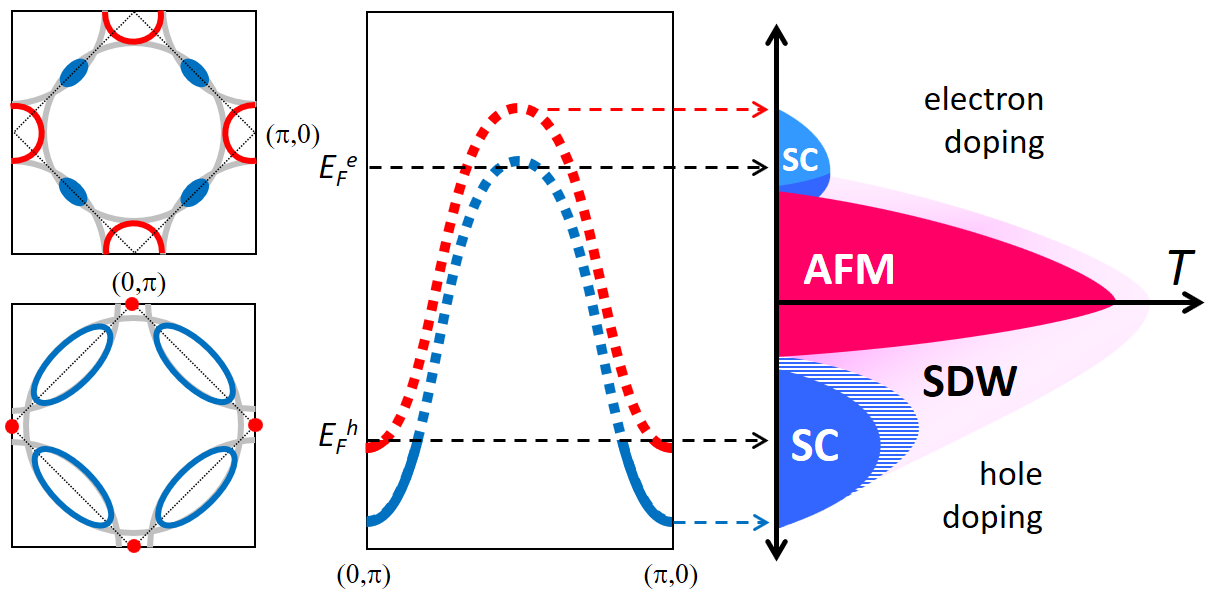}
\caption{Similar projection of the electronic band structure of high-$T_c$ cuprates, represented by the dispersion along the $(\pi,0)-(0,\pi)$ cut of large Brillouin zone (central panel), on their phase diagram (right) \protect\ignorecitefornumbering{\cite{2015_LTP_Kordyuk}}: $T_c$ maxima correspond to proximity of the tops/bottoms of the antiferomagnetically folded bands to the Fermi level; the corresponding Fermi surfaces (blue for hole- and red for electron-like) are shown on the left panels for $E_F^e$ (top) and $E_F^h$ (bottom) Fermi levels.
\label{PhD_CuSC}}
\end{center}
\end{figure*}

Unexpectedly, the recent progress in understanding the mechanisms of pseudogap formation in cuprates (see \cite{2015_LTP_Kordyuk} for review) leads to conclusion that the same LT-$T_c$ correlation takes place also for cuprates (both for the hole- and the electron-doped ones) in the anti-ferromagnetic (AF) Brillouin zone, i.e. assuming that the pseudogap is caused by an AF-like electronic ordering. 

Indeed, it has been shown that while several mechanisms contribute to the pseudogap phenomenon, a short range or slightly incommensurate \cite{2009_PRB_Kordyuk} AF-like ordering stays mostly responsible for the pseudogap openning below $T^*$ \cite{2015_LTP_Kordyuk}. This ordering is most likely a result of VHS nesting \cite{Hashimoto2010}, that is the known mechanism for electronic ordering in "excitonic insulators" \cite{Rossnagel2011}. Some evidence for incommensurate spin density wave (SDW) has been obtained in neutron experiments on YBCO \cite{Haug2010}, while in Refs. \cite{Hashimoto2010, Hashimoto2014} it has been shown that temperature evolution of antinodal ARPES spectrum for Bi-2201 is mostly consistent with a commensurate $(\pi,\pi)$ density wave order.

This means that the superconductivity in cuprates with the highest $T_c$ appears in the AF-ordered "normal" state, the Fermi surfaces of which are shown on the left side of Fig.\;\ref{PhD_CuSC} for the electron (top) and hole (bottom) doped sides of the phase diagram shown on the right. The most representative cut of the electronic bands, taken along the AF Brillouin zone boundary, is shown in center of Fig.\;\ref{PhD_CuSC}. The two bands shown here are the result of the hybridization between original CuO-band and its replica folded into the AF Brillouin zone.

One can see that maximal $T_c$'s are observed when either higher (red) or lower (blue) band are in close vicinity to Lifshitz transition for the hole and electron doped cuprates, respectively, that is intriguingly similar to the Fe-SC case discussed above.

The splitting between these two bands depends of the mechanism of the "AF-like" ordering, that brings us to the old discussion on Slater vs Mott insulators \cite{Friedel1989}. Effective doubling of the unit cell can be described in many ways: as Peierls or spin-Peierls \cite{Jacobs1976} type instability powered by the VHS nesting \cite{Rossnagel2011}, in the extended Hubbard model \cite{Hirsch1984}, and in the $tJ$-model \cite{Spalek1977, Spalek2007} in which the quasiparticles cannot leave the magnetic sublattice in which they were created \cite{Loktev2005}. So, the two old but related questions are arising again: (1) whether it possible to decide between Slater and Mott scenarios based on ARPES data and (2) how this mechanism does affect the electronic structure and, subsequently, the transition to superconducting state. 

In my opinion, for the hole doped cuprates, this question can be clarified looking for the spectral weight which disappears with pseudogap opening below $T^*$ but reappears below $T_c$ \cite{2015_LTP_Kordyuk}. The upper band (in the center panel of Fig.\;\ref{PhD_CuSC}) is not clearly visible in ARPES spectra \cite{Hashimoto2010, Hashimoto2014}, likely because of short range or incommensurate \cite{2009_PRB_Kordyuk} character of the ordering. One can also remind here the complication that comes from the "shadow band" \cite{Aebi1994}, later attributed to structural modulations \cite{2004_PRB_Koitzschb, Mans2006}, and the bilayer splitting \cite{Feng2001, Chuang2001, 2002_PRB_Kordyuk, 2002_PRL_Kordyuk, 2004_PRB_Kordyuk}, that further complicates the spectra of the bilayer cuprates. So, in order to describe the band gap caused by the AF-like ordering one needs a detailed temperature dependent ARPES study of preferably one layer compounds.

The situation is much simpler for the electron doped cuprates \cite{Alff2003}. Despite the chose between Slater and Mott pictures is also discussed here \cite{Das2010, Weber2010}, the ARPES data clearly shows a gap along the magnetic zone boundary \cite{Matsui2005, Park2007} and the Fermi surface like in Fig.\;\ref{PhD_CuSC}) (left-bottom), confirming the AF doubling of the unit cell.

To summarize, the AF-like ordering in cuprates puts them in a row with the Fe-SC's following the empirical correlation that highlights the importance of topological Lifshitz transition for high temperature superconductivity. One may speculate that the presence of small Fermi surfaces is a general mechanism for enhancement of superconductivity that is usually observed at the charge/spin density wave phase boundary in a number of quasi-2D systems, and that it is the size or geometry of these small Fermi-surfaces rather than an enhancement of the density of states that powers this mechanism.   

\begin{figure*}
\begin{center}
\includegraphics[width=1\textwidth]{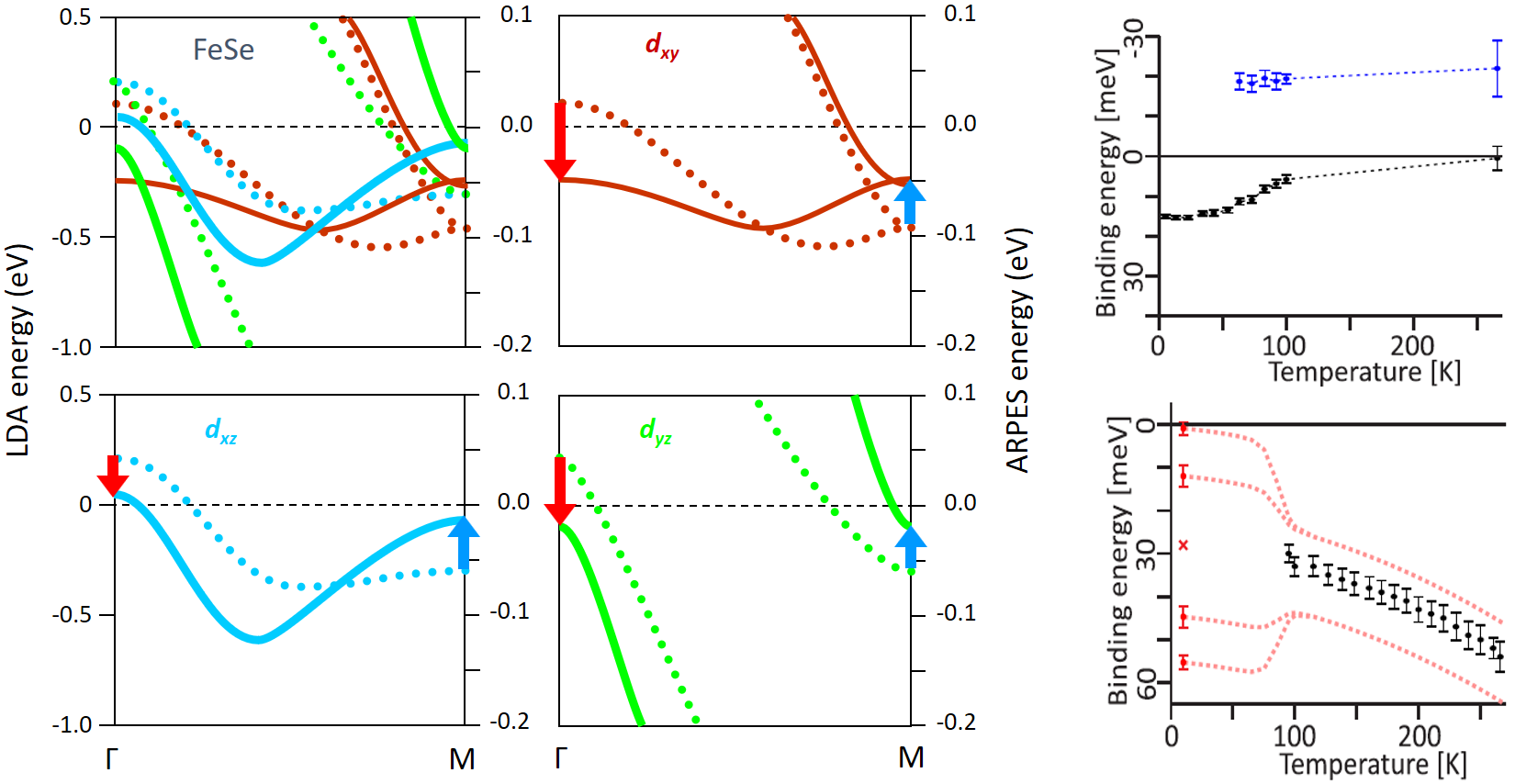}
\caption{The "red-blue shift" of the experimental band structure (the solid lines in four panels on the left) of the FeSe single crystal in comparison to the calculated one (the dotted lines) shown by the red and blue arrows \protect\ignorecitefornumbering{\cite{2016_LTP_Pustovit}} and similar shifts with lowering temperature for the tops of the $d_{xz}$ and $d_{yz}$ bands in the center (top right panel) of the Brillouin zone and for the merging point of these bands in its corner (bottom right panel) \protect\ignorecitefornumbering{\cite{2017_PRB_Kushnirenko}}.
\label{RedBlueShift}}
\end{center}
\end{figure*}

\subsection{Resonant superconductivity?}
\label{RSC}

The question about possible mechanism of $T_c$ enhancement by a proximity to the Lifshitz transition is evidently not straightforward since the very definition of this transition as the "2.5 phase transition" in the terminology of Ehrenfest indicates that one may expect (for 3D system) to observe singularity only in between the 2nd and 3rd derivatives of the thermodynamic potential: $z^{1/2}$ and $z^{-1/2}$ singularities, respectively, where $z = E_F - \varepsilon_\mathrm{VHS}$ is the energy distance of band-edge VHS to the Fermi level \cite{Lifshitz1960}. 

The effect of Lifshitz transition on electronic properties of metals has been reviewed in \cite{Varlamov1989, Blanter1994}. In particular, it was shown that, besides evident appearance of cusps in DOS and consequently in the heat capacity, magnetic susceptibility etc., it leads to a special channel of scattering with a specific energy dependent correction in the electron relaxation time, which is responsible for a giant anomaly in the thermoelectric power and other kinetic characteristics of metals. Although the effect on superconductivity has not been discussed in those reviews, one may expect a similar effect since often the growth of thermopower is correlated with the changes in $T_c$. Moreover, in anisotropic superconductors, the change in $T_c$ due to the change in the Fermi-surface topology becomes stronger and nontrivial \cite{Makarov1975}.

Both Fe-SC's and Cu-SC's are quasi-2D materials with the step-like singularity in DOS at $\varepsilon_\mathrm{VHS}$ instead of $(\varepsilon - \varepsilon_\mathrm{VHS})^{1/2}$ for 3D compounds, so, one may expect stronger effects here. In addition, they are multi-band superconductors, in which these effects can be enhanced. For example, a "superlinear" enhancement (with number of valleys) of effective coupling constant has been predicted within the BCS model for multi-valued semimetals taking into account inter-valley coupling \cite{Pashitskii1978}.

So, one may expect that the microscopic explanation for the observed LT-$T_c$ correlation may be related with some LT-powered "resonant amplifier" of superconducting pairing in a multi-band system. This said, the superconductivity in FeAs-based compounds has been assigned to a Feshbach resonance (also called "shape resonance" \cite{Perali1996}) in the exchange-like interband pairing \cite{Caivano2009}. This mechanism has been further developed in \cite{Innocenti2010, Innocenti2011, Bianconi2013, Bianconi2015} for a number of systems, including the potassium doped p-Terphenyl \cite{Mazziotti2017} with $T_c$ up to 123 K.

Going back to the issue of superconductivity in Fe-SC system, one may subdivide the pairing problem into "glue" and "amplifier". Recently \cite{Kalenyuk2018}, we have added an argument to support the $s^{+-}$ scenario based on the phase sensitive Josephson junction experiment. Taking the spin-fluctuations as a main glue for electron pairs in Fe-SC, one may consider the shape resonance as the mechanism for an entanglement of spin and orbital degrees of freedom. 

To sumarize, one may assume that the proximity of Fermi surface to topological transition is a universal feature for $T_c$ enhancement mechanism. While this mechanism remains to be fully understood, one may notice that the electronic correlations often shifts the electronic bands to optimal for superconductivity positions. In respect to cuprates, there were a number of experimental evidences and theoretical treatments for the effect of pinning of the saddle-point VHS to the Fermi level. The band-edge VHS, discussed here, is formed due to AF ordering and tuned by the pseudogap value. So, the role of the PG in high-$T_c$ story is not just in competition with superconductivity for the phase space, as suggested by Eq.\,\ref{E2}, but also in shifting the upper split band (for the hole doped cuprates) to the Lifshitz transition. Speculating more, this position should be very sensitive to the new order potential and can be pinned to $E_F$ to minimize the electron kinetic energy. 

As for the Fe-SC compounds, it is interesting to note that the observed "red-blue shift"  \cite{2016_LTP_Pustovit, 2017_PRB_Kushnirenko} can be a consequense of similar pinning mechanism.

\section{"Red-blue shift"}

When one compares the band structure of iron based superconductors derived from ARPES experiment with the result of DFT calculations, one can see that it is not "rigid" but distorted by a momentum-dependent shift that acts similarly in all Fe-SC's, shifting the bands up and down in energy: up---in the center of the Brillouin zone and down---in its corners \cite{Yi2009, 2010_PRL_Borisenko, Brouet2013, 2013_JSNM_Kordyuk}, as shown in Fig.\;\ref{RedBlueShift}). Since such a shift persists in all the Fe-SC's and is a sort of natural degree of freedom for the band structure of a multi-band metal with the Luttinger-volume conserved, that results in synchronous change (shrinking, in this case \cite{Coldea2008, Watson2015, Fanfarillo2016}) of the hole and electron Fermi surfaces, it is tempting to give it a special name and, following \cite{2016_LTP_Pustovit, Lochner2017}, I call it the "red-blue shift" here.

Since it is the electron interactions that are missing in DFT calculations, it is also tempting to ascribe such a shift to these interactions, for which several models have been proposed. The Fermi surface shrinking can be a consequence of the strong particle-hole asymmetry of electronic bands assuming a dominant interband scattering \cite{Ortenzi2009} and described by the self-energy corrections due to the exchange of spin fluctuations between hole and electron pockets \cite{Benfatto2011, Fanfarillo2016}, or it can be formulated in terms of $s$-wave Pomeranchuk ordering \cite{Chubukov2016}. On the other hand, one can explain it as a decrease of a band width due to a screening of the nearest neighbor hopping as a result of AF-like ordering \cite{2016_LTP_Pustovit} that, similarly to cuprates, can be considered as a consequence of the confinement of the carriers within the magnetic sub-lattice \cite{Loktev2005}. One should note that all these mechanisms will lead to Lifshitz transition for non-compensated carriers, that requires hole or electron doping to shift from stoichiometry, multiple bands or both.

If the discussed shift is a result of correlations, one may expect its enhancement with lowering temperature \cite{2016_LTP_Pustovit}. Such temperature evolution of the band structure is in agreement with Hall measurements \cite{2011_JPSJ_Evtushinsky} and has been observed recently by ARPES on FeSe crystals \cite{2017_PRB_Kushnirenko}. This results, however, is not confirmed by other experiments \cite{Rhodes2017, Pustovit2017, Pustovit2018}, so, one may conclude that the temperature effect is more complex and requires further research.

\section{Summary}

The electronic band structure of cuprates defines both the spectrum of the spin-fluctuations, which bound electrons in pairs, and the AF-like electronic ordering, which forms the pseudogap state. The band structure of the iron based superconductors is much more complex than of cuprates. The pairing can be due to spin-fluctuations, phonons or both, but why the $T_c$'s are so high is not clear. Nevertheless, there is an empirical correlation between electronic structure and $T_c$: maximal $T_c$ (optimally doped SC) is observed when proximity of the ES to topological Lifshitz transition takes place. This is observed for all Fe-SC’s.

Interestingly, the same correlation holds for Cu-SC (both for hole- and electron-doped ones) in the anti-ferromagnetic Brillouin zone, i.e. assuming that the PG is caused by the AF-like electronic ordering. So, an interplay of the electronic structure with the spin-fluctuations leads to superconductivity that, on one hand, competes with the pseudogap caused by the AF-ordering, but, on the other hand, can be enhanced by the proximity to Lifshitz transition, also caused by this ordering.

The idea of this review was to stress ones more that this correlation is either annoyingly observed by ARPES or predicted by calculations for all the known high-$T_c$ superconductors. This allows to assume that the proximity of Fermi surface to topological transition is a universal feature for a $T_c$ enhancement mechanism. While we are looking for microscopic understanding of this correlation, it can be used to search new high temperature superconductors with much higher transition temperatures.

\begin{acknowledgements}
I acknowledge discussions with A. Abrikosov, L. Alff, A. Avella, A. Bianconi, B. B\"{u}chner, S. V. Borisenko, V. Brouet, A. V. Chubukov, T. Dahm, C. Di Castro, I. Eremin, D. V. Evtushinsky, J. Fink, A. M. Gabovich, M. S. Golden, M. Grilli, D. S. Inosov, I. N. Karnaukhov, A. L. Kasatkin, T. K. Kim, M. M. Korshunov, V. M. Krasnov, A. Lichtenstein, V. M. Loktev, R. Markiewicz, I. V. Morozov, S. G. Ovchinnikov, E. A. Pashitskii, N. M. Plakida, V. M. Pudalov, Yu.\;V. Pustovit, M. V. Sadovskii, D. J. Scalapino, A. V. Semenov, J. Spa{\l}ek, T. Valla, A. A. Varlamov, A. N. Vasiliev, A. N. Yaresko, and V. B. Zabolotnyy. The project was supported by the Ukrainian-German grant from the MES of Ukraine (Project M/20-2017) and by the project No. 6250 by STCU and NAS of Ukraine.
\end{acknowledgements}

\bibliographystyle{PRLlike}
\bibliography{scopt}


\end{document}